\RequirePackage[hyphens]{url}

\documentclass[sigconf]{acmart}

\usepackage{subfigure}
\usepackage{fancyvrb}
\fvset{frame=single,
       samepage=true,
       fontfamily=courier,
       numbers=left,
       numbersep=2pt,
       baselinestretch=0.80,
       fontsize=\small}

\setcopyright{none}
\acmYear{2019}

\renewcommand\footnotetextcopyrightpermission[1]{}
\begin{document}

\title{MementoMap Framework for Flexible and Adaptive Web Archive Profiling}

\author{Sawood Alam}
\orcid{0000-0002-8267-3326}
\affiliation{%
  \institution{Old Dominion University}
  \department{Department of Computer Science}
  \city{Norfolk}
  \state{Virginia}
  \postcode{23529}
  \country{USA}
}
\email{salam@cs.odu.edu}

\author{Michele C. Weigle}
\orcid{0000-0002-2787-7166}
\affiliation{%
  \institution{Old Dominion University}
  \department{Department of Computer Science}
  \city{Norfolk}
  \state{Virginia}
  \postcode{23529}
  \country{USA}
}
\email{mweigle@cs.odu.edu}

\author{Michael L. Nelson}
\orcid{0000-0003-3749-8116}
\affiliation{%
  \institution{Old Dominion University}
  \department{Department of Computer Science}
  \city{Norfolk}
  \state{Virginia}
  \postcode{23529}
  \country{USA}
}
\email{mln@cs.odu.edu}

\author{Fernando Melo}
\affiliation{%
  \institution{FCT: Arquivo.pt}
  \state{Lisbon}
  \country{Portugal}
}
\email{fernando.melo@fccn.pt}

\author{Daniel Bicho}
\affiliation{%
  \institution{FCT: Arquivo.pt}
  \state{Lisbon}
  \country{Portugal}
}
\email{daniel.bicho@fccn.pt}

\author{Daniel Gomes}
\affiliation{%
  \institution{FCT: Arquivo.pt}
  \state{Lisbon}
  \country{Portugal}
}
\email{daniel.gomes@fccn.pt}

\renewcommand{\shortauthors}{S. Alam et al.}

\begin{abstract}
In this work we propose \emph{MementoMap}, a flexible and adaptive framework to efficiently summarize holdings of a web archive.
We described a simple, yet extensible, file format suitable for \emph{MementoMap}.
We used the complete index of the \emph{Arquivo.pt} comprising 5B mementos (archived web pages/files) to understand the nature and shape of its holdings.
We generated \emph{MementoMaps} with varying amount of detail from its \emph{HTML} pages that have an \emph{HTTP} status code of \texttt{200 OK}.
Additionally, we designed a single-pass, memory-efficient, and parallelization-friendly algorithm to compact a large \emph{MementoMap} into a small one and an in-file binary search method for efficient lookup.
We analyzed more than three years of \emph{MemGator} (a Memento aggregator) logs to understand the response behavior of 14 public web archives.
We evaluated \emph{MementoMaps} by measuring their \emph{Accuracy} using 3.3M unique \emph{URIs} from \emph{MemGator} logs.
We found that a \emph{MementoMap} of less than 1.5\% \emph{Relative Cost} (as compared to the comprehensive listing of all the unique original \emph{URIs}) can correctly identify the presence or absence of 60\% of the lookup \emph{URIs} in the corresponding archive while maintaining 100\% \emph{Recall} (i.e., zero false negatives).
\end{abstract}


\maketitle

\section{Introduction}

Old Dominion University (ODU) runs \emph{MemGator}~\cite{memgator} as a service to power many of our tools and services such as Mink~\cite{minkjcdl14}, CarbonDate~\cite{carbondate}, WAIL~\cite{wail:jcdl17}, ICanHazMemento~\cite{icanhaz}, and MementoDamage~\cite{damage}.
We released \emph{MemGator}~\cite{memgator:gh} as an open-source tool for users to run locally to avoid generating too much traffic on a central aggregator service.
Our service receives three aggregation lookup requests per minute on average.
Due to this low traffic we do not yet use any prediction-based Memento routing or caching.
We recently analyzed over three years of our \emph{MemGator} logs and found that it has served about 5.2M requests so far.
These lookups were broadcasted to 14 different upstream archives for a total of 61.8M requests.
Only 5.44\% of these requests had a hit, while the remaining 93.56\% were either a miss or an error as shown in Table~\ref{tab:mg-log-in-archives}.
If only there was a way to know a summary of the holdings of these web archives, we could have avoided many wasted upstream requests and had an overall better response time for clients.

\emph{MementoMap} is a framework for profiling web archives and expressing their holdings in an adaptive and flexible way to easily scale.
It is inspired by the simplicity of the widely used \texttt{robots.txt} and \texttt{sitemap.xml} formats, but for a purpose other than search engine optimization.
An example \emph{MementoMap} is illustrated in Figure~\ref{code:ukvs} in the format we propose.
\emph{MementoMap} allows wildcard-based partial \emph{URI Keys} to enable flexibility in how detailed or concise one wants it to be depending on use cases, full or partial knowledge about the archive's holdings, and available resources.
This can either be generated by the archives themselves or by a third party based on their external observations.
We propose the ``\texttt{mementomap}'' \emph{well-known URI} suffix~\cite{rfc8615} and the ``\texttt{mementomap}'' \emph{link relation} for its dissemination and discovery.

We used the complete index of \emph{Arquivo.pt} (the Portuguese Web Archive), spanning over 27 years, and more than three years of \emph{MemGator} logs for evaluation.
We found that a summarized \emph{MementoMap} of less than 1.5\% \emph{Relative Cost} (as compared to the comprehensive listing of all the unique original \emph{URIs}) can correctly identify the presence or absence of 60\% of the lookup \emph{URIs} in \emph{Arquivo.pt} without any false negatives (i.e., 100\% \emph{Recall}).
We have open-sourced our implementation~\cite{mementomap:gh} under the \emph{MIT} license.
This paper is an expanded version of a conference paper~\cite{mementomap:jcdl19}.

\section{Background}

The Internet Archive (IA)\footnote{\url{https://archive.org/}} is the first, largest, and most resourceful web archive with over 700B mementos (timestamped archived copies of web pages and files) as of January 21, 2019\footnote{\url{https://twitter.com/brewster_kahle/status/1087515601717800960}}.
However, it is also the softest target for censorship and denial of service attacks~\cite{iaddos}.
It continues to be blocked in China~\cite{iablock:china} and Russia~\cite{iablock:russia} for an extended period of time and has been blocked temporarily in many other countries such as India and Jordan~\cite{iablock:india-ia,iablock:jordan}.
As a result, many web archiving related tools are increasingly adding support for Memento aggregators to consolidate archived resources from more than one web archive of varying scale to avoid single point of failure.\looseness=-1

The Memento framework~\cite{memento:rfc} defines uniform APIs for \emph{TimeMap} and \emph{TimeGate} endpoints to enable cross-archive communication.
A \emph{TimeMap} is a list of all mementos of an original URI (or \emph{URI-R}) and a \emph{TimeGate} is a gateway to resolve to the closest memento of a \emph{URI-R} w.r.t. a given \emph{Datetime} and redirect to a Memento URI (or \emph{URI-M}).
With out-of-the-box Memento support in major archival tools and replay systems, many web archives have adopted the protocol.
To avoid the need of every tool being configured and periodically updated to poll results from an ever-changing list of many known web archives, Memento aggregators were created to act like a single consolidated web archive to users and tools.
Los Alamos National Laboratory's (LANL) \emph{Time Travel}~\footnote{\url{http://timetravel.mementoweb.org/}} service is one such well-known aggregator that powers many tools and services.
\emph{MemGator} is our open-source Memento aggregator implementation that can be used locally as a CLI tool or run as a service for a drop-in replacement of the \emph{Time Travel} service.

\emph{CDX (Capture inDeX)}~\cite{cdxformat} is a \emph{CSV}-like text file-based index format that has traditionally been used by the IA and was one of the primarily supported index formats of OpenWayback\footnote{\url{https://github.com/iipc/openwayback}}.
It is very rigid in nature and has a predefined list of fields that are not extendable.
While working on this paper, we encountered a consequence of its limitations when we realized that the \emph{MIME-Type} field was reused to record a different metadata to identify whether a record is a \emph{revisit}.
As a result the actual \emph{MIME-Type} of the record would not be known without finding another entry in the index which the record is a \emph{revisit} of.
\emph{CDXJ}~\cite{cdxjspec} is an evolution of the classic \emph{CDX} format.
In this file format, lookup key fields (\emph{URI-R} and \emph{Datetime}) are placed at the beginning of each line which is followed by a single-line compact JSON~\cite{rfc4627} block that holds other fields that can vary in number and be extended as needed.
Both of these formats are sort-friendly to enable binary search on file when performing lookups.
The latter format is primarily used by archival replay systems including PyWB\footnote{\url{https://github.com/webrecorder/pywb}} and our InterPlanetary Wayback (IPWB)~\cite{ipwb-tpdl2016}.

\definecolor{softpurple}{RGB}{166, 77, 255}
\definecolor{softgreen}{RGB}{0, 128, 96}
\definecolor{softred}{RGB}{255, 77, 148}
\definecolor{softbrown}{RGB}{210, 121, 121}

\begin{SaveVerbatim}[commandchars=^\{\}]{VerbEnv}
^textcolor{softpurple}{http://}^textbf{^textcolor{softgreen}{(com,cnn}^textcolor{softred}{)/*}}
^textcolor{softpurple}{http://}^textbf{^textcolor{softgreen}{(com,cnn,cdn}^textcolor{softred}{)/img/logos/logo.png}}^textcolor{softbrown}{?h=20&w=30}
^textcolor{softpurple}{http://}^textbf{^textcolor{softgreen}{(com,nytimes}^textcolor{softred}{)/2018/10/*}}
^textcolor{softpurple}{http://}^textbf{^textcolor{softgreen}{(com,nytimes}^textcolor{softred}{)/2018/11/*}}
^textcolor{softpurple}{http://}^textbf{^textcolor{softgreen}{(edu,odu,cs,ws-dl}^textcolor{softred}{)/}}
^textcolor{softpurple}{http://}^textbf{^textcolor{softgreen}{(org,arxiv}^textcolor{softred}{)/*}}
^textcolor{softpurple}{http://}^textbf{^textcolor{softgreen}{(org,arxiv}^textcolor{softred}{)/pdf/*}}
^textcolor{softpurple}{http://}^textbf{^textcolor{softgreen}{(uk,bl,*}}
^textcolor{softpurple}{http://}^textbf{^textcolor{softgreen}{(uk,co,bbc}^textcolor{softred}{)/news/world}}^textcolor{softbrown}{?lang=ar}
^textcolor{softpurple}{http://}^textbf{^textcolor{softgreen}{(uk,co,bbc}^textcolor{softred}{)/news/world}}^textcolor{softbrown}{?lang=en}
^textcolor{softpurple}{http://}^textbf{^textcolor{softgreen}{(uk,gov,*}^textcolor{softred}{)/}}
\end{SaveVerbatim}

\begin{figure}
\subfigure[A sample list of sorted \emph{SURTs}.
           Different colors signify \emph{Scheme}, \emph{Host}, \emph{Path}, and \emph{Query} segments.
           The ``\texttt{https://(}'' prefix is common in all \emph{SURTs}, hence removed in practice.]{
  \label{img:surt-list}
  \begin{minipage}{\linewidth}
    \UseVerbatim[numbers=none]{VerbEnv}
  \end{minipage}
}
\subfigure[A visual representation of \emph{SURTs} as a tree.
           Different colored regions signify \emph{Scheme}, \emph{Host}, \emph{Path}, and \emph{Query} segments.
           Each node of the tree contains a token and each edge denotes the separator of the corresponding segment.
           Dotted lines indicate transition from one segment to the next.
           Dotted triangles with a wildcard character ``*'' denote a sub-tree.
           Trailing slashes are removed from this representation.
           Labels on the right hand side (i.e., \texttt{S}, \texttt{H0}--\texttt{Hn}, \texttt{P0}--\texttt{Pn}, and \texttt{Q}) denote corresponding level/depth in each segment.]{
  \label{img:surt-tree}
  \includegraphics[width=\linewidth]{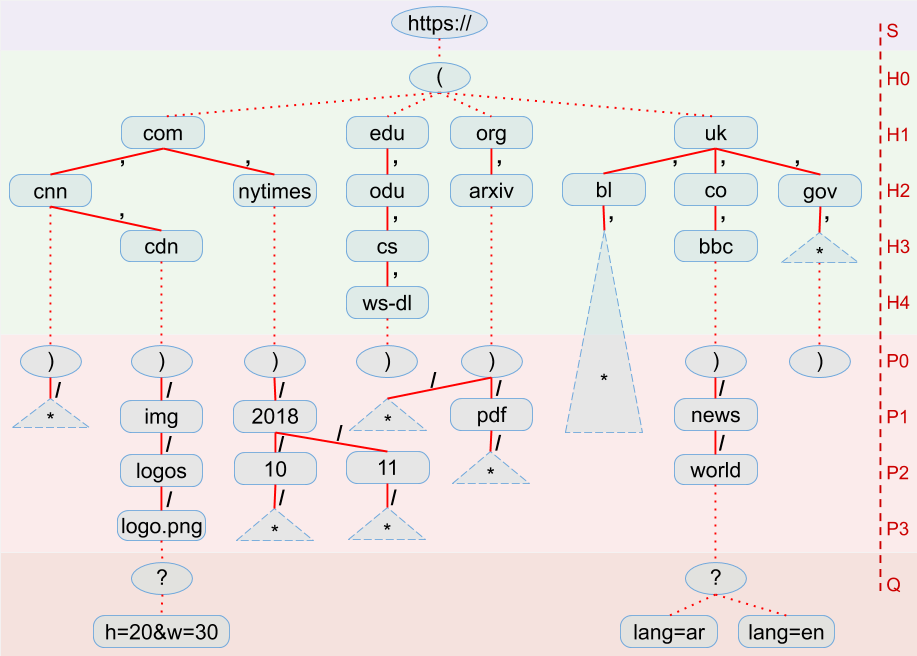}
}
\caption{Illustration of \emph{SURTs} with wildcard.}
\label{img:surt-example}
\end{figure}

\emph{SURT (Sort-friendly URI Reordering Transform)}~\cite{surt} is used to canonicalize \emph{URIs} and place together related \emph{URIs} when sorted, which is important for efficient indexing.
In a traditional URI the hostname parts are organized differently than paths.
In the hostname section, the root of the Domain Name System (\emph{DNS}) chain (i.e., the Top Level Domain, or \emph{TLD}) comes at the end towards the right hand side while registered domain name portion and subdomain sections are placed towards the left hand side.
In contrast, in the path section, the root path comes first followed by deeper nodes of the path tree towards the right side.
As a consequence, if a list of three domain names \texttt{example.com}, \texttt{foo.example.com}, and \texttt{example.net} are sorted, the latter with a different \emph{TLD} will sit in between the other two.
As opposed to this the \emph{SURT} of ``\texttt{Www.Foo.Example.COM/a/b?x=y\&c=d}'' converts it to become \\ ``\texttt{com,example,foo)/a/b?c=d\&x=y}'', which changes the domain\\ name with lower case letters, removes the ``\texttt{www}'' subdomain, reverses the order of hostname segments, and sorts query parameters.
\emph{SURTs} are commonly used in archival index files and many other places where a URI is used as a lookup field, including \emph{MementoMap}.
Figure~\ref{img:surt-list} illustrates a sample of sorted \emph{SURTs} and highlights different segments.
We have extended \emph{SURTs} to support wildcard to allow grouping of \emph{URI Keys} with the same prefix and roll them up into a single key.
A visual representation of these \emph{SURTs} is illustrated in Figure~\ref{img:surt-tree} in the form of a tree that segregates layers of \emph{Scheme}, \emph{Host}, \emph{Path}, and \emph{Query}.
It further annotates various depths of \emph{Host} and \emph{Path} segments as \emph{H0, H1, H2\ldots} and \emph{P0, P1, P2\ldots} that will be useful in understanding some terminologies used later in this paper.
\emph{SURTs} also allow credentials and port numbers, but we omitted them from the illustration for brevity.
It is worth noting that the scheme portion is common in all \emph{HTTP/HTTPS} URIs and has no informational value, hence the ``\texttt{https://(}'' prefix is often omitted.\looseness=-1

\emph{UKVS (Unified Key Value Store)}~\cite{ukvs} is an evolving file format proposal that is a contribution of this \emph{MementoMap} work.
It is an evolution of the \emph{CDXJ} format that we earlier proposed to be used by \emph{Archive Profiles}~\cite{alam-ijdl16-profiling}.
This format extends \emph{SURT} with wildcard support and improves various other aspects to simplify it and eliminate some limitations of our prior proposal (such as not being able to express blacklists or lack of support to merge two profiles generated with different profiling policies).
We generalized the format to be more inclusive and flexible after we realized its utility in many web archiving related use cases (such as indexing, replay access control list (ACL), fixity blocks~\cite{fixityblock}, and extended \emph{TimeMaps}) and many other places such as extended server logging.

\begin{figure}
\begin{Verbatim}[commandchars=\\^?]
\textit^!context ["https://git.io/mementomap"]?
\textit^!id      {uri: "https://archive.example.org/"}?
\textit^!fields  {keys: ["surt"], values: ["frequency"]}?
\textit^!meta    {name: "Example Archive", year: 1996}?
\textit^!meta    {type: "MementoMap"}?
\textit^!meta    {updated_at: "2018-09-03T13:27:52Z"}?
\textbf^*?                   \textbf^54321/20000?
\textbf^com,*?               \textbf^10000+?
\textbf^org,arxiv)/?         \textbf^100?
\textbf^org,arxiv)/*?        \textbf^2500~/900?
\textbf^org,arxiv)/pdf/*?    \textbf^0?
\textbf^uk,co,bbc)/images/*? \textbf^300+/20-?
\end{Verbatim}
\caption{A sample \emph{MementoMap} in \emph{UKVS} format.
         \textnormal{Lines beginning with a ``\texttt{!}'' denote headers.
                     Lines in bold text are data entries.
                     The ``\texttt{!fields}'' line describes keys and values in data columns in order.
                     The ``\texttt{frequency}'' column of the data section is formatted as ``\texttt{[URI-M Count]/[URI-R Count]}''.
                     Optional suffix characters (i.e., \texttt{+}, \texttt{-}, and \texttt{\textasciitilde}) with numbers denote approximate values.
                     A ``\texttt{0}'' value is a way to represent blacklists, potentially, for more specific sub-trees.}
         }
\label{code:ukvs}
\end{figure}

Figure~\ref{code:ukvs} illustrates a sample \emph{MementoMap} file that starts with some metadata headers.
Header lines are prefixed with ``\texttt{!}'' to ensure they are separated from data lines and surfaced on top when the file is sorted.
The ``\texttt{!fields}'' header tells that the first column is a \emph{SURT} and is used as a lookup key (there can be more than one key column such as \emph{Datetime} or \emph{Language}) which is followed by a value column that holds ``\texttt{frequency}'' information.
Each data line can optionally also contain a single-line \emph{JSON} block, which is not illustrated here for the sake of simplicity.
The \texttt{frequency} column is formatted as ``\texttt{[URI-M Count]/[URI-R Count]}'' where both counts are optional and the separator is also optional if only the \texttt{URI-M Count} is present (in this paper we only used this latter option).
Additionally, these counts can have an optional suffix character \texttt{+}, \texttt{-}, or \texttt{\textasciitilde} to express that the numbers are not exact and represent a lower bound, an upper bound, and a rough estimate respectively.
The first data line in the example means there are a total of exactly 54,321 mementos (\emph{URI-Ms}) of exactly 20,000 \emph{URI-Rs} in the archive and the next line suggests that there are at least 10,000 mementos from the ``\texttt{.com}'' \emph{TLD}.
The next two lines suggest that there are 100 mementos of the \texttt{arxiv.org} homepage and many more captures of pages with deeper paths.
However, the next line illustrates an exclusion of a sub-tree by being more specific under \texttt{/pdf/*} that has zero mementos (this is illustrated in Figure~\ref{img:surt-tree} as well).

\begin{table}
  \caption{Top \emph{Arquivo.pt} \emph{TLDs}.}
  \label{tab:top-tld}
  \begin{tabular}{l | r r}
    \toprule
    \textbf{TLD}   & \textbf{URI-R\%} & \textbf{URI-M\%} \\
    \midrule
    \texttt{.pt}   &           61.422 &           68.266 \\
    \texttt{.com}  &           19.610 &           19.643 \\
    \texttt{.eu}   &            8.665 &            4.262 \\
    \texttt{.net}  &            1.973 &            1.829 \\
    \texttt{.org}  &            1.790 &            1.263 \\
    \texttt{.de}   &            0.635 &            0.343 \\
    \texttt{.br}   &            0.617 &            0.470 \\
    \texttt{.uk}   &            0.449 &            0.260 \\
    \texttt{.fr}   &            0.347 &            0.173 \\
    \texttt{.nl}   &            0.274 &            0.131 \\
    \texttt{.mz}   &            0.236 &            0.414 \\
    \texttt{.pl}   &            0.226 &            0.104 \\
    \texttt{.io}   &            0.223 &            0.208 \\
    \texttt{.edu}  &            0.201 &            0.096 \\
    \texttt{.es}   &            0.200 &            0.126 \\
    \texttt{.it}   &            0.198 &            0.109 \\
    \texttt{.cv}   &            0.198 &            0.335 \\
    \texttt{.ru}   &            0.196 &            0.203 \\
    \texttt{.ao}   &            0.156 &            0.295 \\
    \texttt{.us}   &            0.142 &            0.102 \\
    \texttt{.cz}   &            0.117 &            0.057 \\
    \texttt{.info} &            0.113 &            0.160 \\
    \midrule
    IP Addresses   &            0.070 &            0.050 \\
    Other TLDs     &            1.941 &            1.149 \\
    \bottomrule
  \end{tabular}
\end{table}

\begin{table}
  \caption{\emph{Arquivo.pt} index statistics.}
  \label{tab:pt-cdxj-stats}
  \begin{tabular}{l | r}
    \toprule
    \textbf{Attributes}  & \textbf{Values} \\
    \midrule
    CDXJ files           &              70 \\
    Total file size      &            1.8T \\
    Compressed file size &            262G \\
    Temporal coverage    &      1992--2018 \\
    CDXJ lines           &            5.0B \\
    Mementos (URI-Ms)    &            4.9B \\
    Unique URI-Rs        &            2.0B \\
    Unique HxPx keys     &            1.1B \\
    Unique hosts         &            5.8M \\
    Unique IP addresses  &             15K \\
    \bottomrule
  \end{tabular}
\end{table}

\emph{Arquivo.pt}~\cite{pwa} was founded in 2008 with the aim to preserve web content of interest to the Portuguese community, but not limited to just the \texttt{.pt} \emph{TLD} (as shown in Table~\ref{tab:top-tld}).
It has since archived about 5B mementos of which some data was donated to it by other archives, including IA, explaining why its temporal spread extends back before the \emph{Arquivo.pt}'s founding date.
We analyzed 1.8T of \emph{Arquivo.pt}'s complete \emph{CDXJ} index in production.
A brief summary of the dataset is shown in Table~\ref{tab:pt-cdxj-stats}.
We used it along with ODU's \emph{MemGator} server logs to evaluate this work.

\section{Related Work}

Query routing is a rigorously researched topic in various fields including, networked databases, meta-searching, and search aggregation~\cite{starts,metasearch:survey}.
However, archive profiling and Memento lookup routing is a niche field that is not explored by many researchers beyond a small community.

Sanderson et al. created comprehensive content-based profiles~\cite{profurir,mementointegration} of various \emph{International Internet Preservation Consortium (IIPC)} member archives by collecting their \emph{CDX} files and extracting URI-Rs from them.
This approach gave them complete knowledge of the holdings in each participating archive, hence they can route queries precisely to archives that have any mementos for the given \emph{URI-R}.
This approach yielded no false positives or false negatives (i.e., 100\% \emph{Accuracy}) while the \emph{CDX} files were fresh, but they would go stale very quickly.
It is a resource and time intensive task to generate such profiles and some archives may be unwilling or unable to provide their \emph{CDX} files.
Such profiles are so large in size (typically, a few billion \emph{URI-R} keys) that they require special infrastructure to support fast lookup.
Acquiring fresh \emph{CDX} files from various archives and updating these profiles regularly is not easy.

In contrast, AlSum et al. explored a minimal form of archive profiling using only the \emph{TLDs} and \emph{Content-Language}~\cite{proftldlangtpdl,proftldlang}.
They created profiles of 15 public archives using access logs of those archives (if available) and fulltext search queries.
They found that by sending requests to only the top three archives matching the criteria for the lookup URI based on their profile, they can discover about 96\% of \emph{TimeMaps}.
When they excluded IA from the list and performed the same experiment on the remaining archives, they were able to discover about 65\% of \emph{TimeMaps} using the remaining top three archives.
Excluding IA was an important aspect of evaluation as its dominance can cause bias in results.
This exclusion experiment also showed the importance of smaller archives and the impact of their holdings collectively.
This minimal approach had many false positives, but no false negatives.

Bornand et al. implemented a different approach for Memento routing by building binary classifiers from LANL's Time Travel aggregator cache data~\cite{routeclass}.
They analyzed responses from various archives in the aggregator's cache over a period of time to learn about the holdings of different archives.
They reported a 77\% reduction in the number of requests and a 42\% reduction in response time while maintaining 85\% \emph{Recall}.
These approaches can be categorized as usage-based profiling in which access logs or caches are used to observe what people were looking for in archives and which of those lookups had a hit or miss in the past.
While usage-based profiling can be useful for Memento lookup routing, it may not give the real picture of archives' holdings, producing both false negatives and false positives\footnote{\url{https://groups.google.com/forum/\#!topic/memento-dev/YE4rt6L5ICg}}.

We found that traffic from \emph{MemGator} requested less than 0.003\% of the archived resources in \emph{Arquivo.pt}.
There is a need for content-based archive profiling which can express what is present in archives, irrespective of whether or not it is being looked for.

In previous work~\cite{arcproftpdl15,alam-ijdl16-profiling}, we explored the middle ground where archive profiles are neither as minimal as storing just the \emph{TLD} (which results in many false positives) nor as detailed as collecting every URI-R present in every archive (which goes stale very quickly and is difficult to maintain).
We first defined various profiling policies, summarized \emph{CDX} files according to those policies, evaluated associated costs and benefits, and prepared gold standard datasets~\cite{arcproftpdl15,alam-ijdl16-profiling}.
In our experiments, we correctly identified about 78\% of the URIs that were or were not present in the archive with less than 1\% relative cost as compared to the complete knowledge profile and identified 94\% URIs with less than 10\% relative cost without any false negatives.
Based on the archive profiling framework we established, we further investigated the possibility of content-based profiling by issuing fulltext search queries (when available) and observing returned results~\cite{arcproftpdl16} if access to the \emph{CDX} data is not possible.
We were able to make routing decisions of 80\% of the requests correctly while maintaining about 90\% \emph{Recall} by discovering only 10\% of the archive holdings and generating a profile that costs less than 1\% of the complete knowledge profile.
\emph{MementoMap} is a continuation of this effort to make it more flexible and portable by eliminating the need for rigid profiling policies we defined earlier~\cite{arcproftpdl15,alam-ijdl16-profiling} (which are still good for baseline evaluation purposes) and replacing them with an adaptive approach in which the level of detail is dynamically controlled with a number of parameters.\looseness=-1

\section{Methodology}

\begin{figure}
\begin{Verbatim}[commandchars=^\{\}]
func ^textbf{host_keys}(surt)
  s = surt.split(")")[0].split(",", MAXHOSTDEPTH)
  return [s[:i].join(",") for i in 1..len(s)]

func ^textbf{path_keys}(surt)
  s = surt.split("?")[0].split("/", MAXPATHDEPTH)
  return [s[:i].join("/") for i in 1..len(s)]

func ^textbf{compact}(imap, omap, opts)
  htrail = [None] * MAXHOSTDEPTH
  ptrail = [None] * MAXPATHDEPTH
  for line in imap
    key, freq, *_ = line.split()
    k = host_keys(key)
    for i in range(len(k))
      if htrail[i] == k[i] ^textcolor{gray}{^textit{# Existing branch}}
        htrail[i][1] += freq
      else ^textcolor{gray}{^textit{# New branch}}
        for j in range(i, MAXHOSTDEPTH)
          if rollup_threshold_reached
            omap.seek(htrail[j][3]) ^textcolor{gray}{^textit{# Move back}}
            omap.write(htrail[j][:1].join(",* "))
            reset_remaining_trail(ptrail, 0)
        reset_remaining_trail(htrail, i)
      if !htrail[i] ^textcolor{gray}{^textit{# New tree node}}
        htrail[i] = [k[i], freq, 0, omap.tell()]
        htrail[i-1][2]++ ^textcolor{gray}{^textit{# Incr parent's children}}
    ^textcolor{gray}{^textit{# Repeat similar logic for path segment}}
    omap.write(line)
    omap.truncate() ^textcolor{gray}{^textit{# Clear any rollup residue}}

func ^textbf{lookp_keys}(uri)
  key = surtify(uri).split("?")[0].strip("/")
  keys = [key]
  while "," in key
    keys.uppend(sub("(.+[,/]).+$", "\1*", key))
    key = sub("(.+)[,/].+$", "\1", key)
  return keys

func ^textbf{bin_search}(mmap, key)
  surtk, freq, *_ = mmap.readline().split()
  if key == surtk ^textcolor{gray}{^textit{# First line matched}}
    return [surtk, freq]
  left = 0
  mmap.seek(0, 2) ^textcolor{gray}{^textit{# Go to the EOF}}
  right = mmap.tell()
  while (right - left > 1)
    mid = (right + left) / 2
    mmap.seek(mid)
    mmap.readline() ^textcolor{gray}{^textit{# Skip partial line}}
    surtk, freq, *_ = mmap.readline().split()
    if key == surtk
      return [surtk, freq]
    elif key > surtk
      left = mid
    else:
      right = mid

func ^textbf{lookup}(mmap, uri)
  for key in lookp_keys(uri)
    result = bin_search(mmap, key)
    if result
      return [key, result]
\end{Verbatim}
\caption{\emph{MementoMap} Compaction and Lookup procedures.
         \textnormal{These pseudo-code illustrations are not in any specific language.
                     Actual implementation is more elaborated.}
        }
\label{code:compact-lookup}
\end{figure}

Generating a \emph{MementoMap} begins by scanning \emph{CDX/CDXJ} files, performing fulltext search, filtering access logs, or any other means to identify what \emph{URIs} an archive holds (or does not hold).
These \emph{URIs} are then converted to \emph{SURTs} (if not already) and their query section is stripped off.
We call these partial \emph{SURTs} as \emph{HxPx URI Keys} (which means a \emph{URI Key} that has all the host and path parts, but no query parameters).
Previously, we found that removing query parameters from these \emph{SURTs} reduces the file size and the number of unique \emph{URI Keys} significantly without any significant loss in the lookup \emph{Accuracy}~\cite{alam-ijdl16-profiling}.
We then create a text file with its first column containing \emph{HxPx Keys} and the second column as their respective \emph{Frequencies}.
The \emph{frequency} column in its simplest form can be the count of each \emph{HxPx Key}, but it can be made more expressive as illustrated in the data section of Figure~\ref{code:ukvs}.
Finally, necessary metadata is added and the file is sorted as the baseline \emph{MementoMap}.\looseness=-1

In order to make a less detailed \emph{MementoMap} (which is desired for efficient dissemination and long-lasting freshness at the cost of increased false positives), we pass a detailed \emph{MementoMap} through a compaction procedure which yields a summarized output that contains fewer lookup keys by rolling sub-trees with many children nodes up and replacing them with corresponding wildcard keys.
Our compaction algorithm is illustrated with pseudo-code in Figure~\ref{code:compact-lookup}.
As opposed to an in-memory tree building (which will not scale), it is a single-pass procedure with minimal memory requirements and does not need any special hardware to process a \emph{MementoMap} of any size.
We leverage the fact that the input \emph{MementoMap} is sorted, hence, we can easily detect at what depth of host or path segments a branch differed from the previous line.
We keep track of the most recent state of host and path keys at each depth (up to \texttt{MAXHOSTDEPTH} and \texttt{MAXPATHDEPTH}), their corresponding cumulative frequencies, how many children nodes each of them have seen so far, and the byte position of the output file when these keys were seen the first time.
Each time we encounter a new branch at any depth, we check to see if a roll up action is applicable at that depth or further down in the existing tree based on the most recent states and the compaction parameters supplied.
If so, we move the write pointer in the output file back to the position where the corresponding key was observed first, then we reset the state of all the deeper depths and update them with the current state.
As a consequence of this progressive processing, the trailing part of the output file is overwritten many times.
The input file does not have to be the baseline \emph{MementoMap}, any \emph{MementoMap} can be supplied as input with fresh compaction parameters to attempt to further compact it.
Our algorithm is parallel processing-friendly if the input data is partitioned strategically (e.g., processing each \emph{TLD}'s records on separate machines and combining all compacted output files).
It is worth noting that sub-trees of the path section are neither independent trees nor have a single root node (as shown in Figure~\ref{img:surt-tree}), as a result, certain implementation details can be more complex than a simple tree pruning algorithm.

The algorithm for lookup in a \emph{MementoMap} is also illustrated in Figure~\ref{code:compact-lookup}.
Given a \emph{URI}, we first generate all possible lookup keys, in which all keys but the longest one have a wildcard suffix (e.g, ``\texttt{Www.Example.COM/a/b?x=y\&c=d}'' yields ``\texttt{com,example)/a/b}'',\\ ``\texttt{com,example)/a/b/*}'', ``\texttt{com,example)/a/*}'', ``\texttt{com,example)/*}'', and ``\texttt{com,*}'' as lookup keys).
We then perform a binary search in the \emph{MementoMap} with lookup keys in decreasing specificity until we find a match or all the keys are exhausted.
In case of a match, we return the matched lookup key and corresponding frequency results.\looseness=-1

For dissemination and discovery of \emph{MementoMaps} we propose that web archives make their \emph{MementoMap} available at the \emph{well-known URI}~\cite{rfc8615} ``\texttt{/.well-known/mementomap}'' under their domain names.
Alternatively, a custom \emph{URI} can be advertised using the ``\texttt{mementomap}'' \emph{link relation} (or ``\texttt{rel}'') in an \emph{HTTP} \emph{Link} header or \emph{HTML} \emph{<link>} element.
Third parties hosting \emph{MementoMaps} of other archives can use the ``\emph{anchor}'' attribute of the \emph{Link} header to advertise a different context.
Moreover, \emph{MementoMaps} are self-descriptive as they contain sufficient metadata in their headers to establish a relationship with their corresponding archives.
\emph{MementoMaps} support pagination that can be discovered after retrieving the primary \emph{MementoMap} from a \emph{well-known URI} or by any other means.

\section{Evaluation}

For evaluation we used the complete index of \emph{Arquivo.pt}, complete logs of our \emph{MemGator} service, and generated \emph{MementoMaps}.
We first examine logs, then describe holdings of \emph{Arquivo.pt} in detail, and finally measure the effectiveness of various \emph{MementoMaps}.

\subsection{Archived vs. Accessed Resources}

\begin{table}
  \caption{\emph{MemGator} log responses from various archives.
          \textnormal{Data ranges from 2015-10-25 to 2019-01-16.}
          }
  \label{tab:mg-log-in-archives}
  \begin{tabular}{l | r r r r r}
    \toprule
    \textbf{Archive}    & \textbf{Request} & \textbf{Hit\%} & \textbf{Miss\%} & \textbf{Err\%} & \textbf{Sleep} \\
    \midrule
    Internet Archive    &        4,723,880 &          35.76 &           63.68 &           0.56 &          1,594 \\
    Archive-It          &        5,011,385 &           9.14 &           90.38 &           0.48 &          1,556 \\
    Archive Today       &        5,151,720 &           8.44 &           88.96 &           2.60 &          1,920 \\
    Library of Congress &        4,862,458 &           4.77 &           94.31 &           0.92 &          2,705 \\
    Arquivo.pt          &        4,300,221 &           3.35 &           96.29 &           0.36 &          1,153 \\
    Icelandic           &        5,126,706 &           2.22 &           97.14 &           0.64 &          3,143 \\
    Stanford            &        5,178,835 &           1.54 &           98.02 &           0.43 &          1,482 \\
    UK Web Archive      &        5,113,984 &           1.49 &           86.30 &          12.20 &          2,779 \\
    Perma               &        4,116,099 &           1.32 &           98.67 &           0.01 &             46 \\
    PRONI               &        5,165,805 &           0.75 &           98.72 &           0.54 &          1,608 \\
    UK Parliament       &        5,181,991 &           0.63 &           98.85 &           0.52 &          1,542 \\
    NRS                 &        2,683,311 &           0.21 &           99.77 &           0.01 &             46 \\
    UK National         &        5,178,184 &           0.10 &           99.45 &           0.45 &          1,457 \\
    PastPages           &           22,058 &           0.00 &           62.90 &          37.10 &              0 \\
    \midrule
    \textbf{All}        &       61,816,637 &           5.44 &           92.92 &           1.64 &         21,031 \\
    \bottomrule
  \end{tabular}
\end{table}

We analyzed over three years of our \emph{MemGator} logs containing records about 14 different web archives.
In its lifetime it has served a total of 5,241,771 requests for 3,282,155 unique \emph{URIs}.
Table~\ref{tab:mg-log-in-archives} shows the the summary of our log analysis in which IA has over 35\% hit rate, and every other archive is below 10\% (down to zero) in decreasing order of hit rate.
\emph{Arquivo.pt} is showing a 3.35\% hit rate, so we cross checked it with the full index and found that there are only 1.64\% unique \emph{URIs} from the \emph{MemGator} logs that are present in \emph{Arquivo.pt} (note that the \emph{CDX} data even includes recent mementos that would have generated a miss prior to them being archived).
The difference in these numbers is perhaps as a result of some archived \emph{URIs} being looked for more frequently.
This low percentage of overlap in access logs and archive indexes conforms to our earlier findings~\cite{alam-ijdl16-profiling}.
The table shows an overall 93\% miss rate, which is all wasted traffic and delayed response time.
Identifying sources of such a large miss rate can save resources and time significantly, which is the primary motivation of this work.

There are some other notable entries in Table~\ref{tab:mg-log-in-archives} such as low number of requests to PastPages which was excluded from being polled in the early days due to its zero hit rate and high error rate.
NRS (National Records of Scotland) is a new addition to the list, hence it shows a low number of requests.
The high error rate of the UK Web Archive was primarily caused by a bug in the Go language (used to develop \emph{MemGator}) that was not cleaning idle TCP connections that were already closed by the application.
As a result, UKWA's firewall was seeing an ever increasing number of open, but idle connections, hence dropping packets after a hard limit of 20 concurrent connections per host.
This has since been fixed after the release of the Go language version 1.7.
We have later introduced an automatic dormant feature that puts an upstream archive to sleep for a configurable amount of time after a set number of successive errors.

\begin{figure}
  \includegraphics[width=\linewidth]{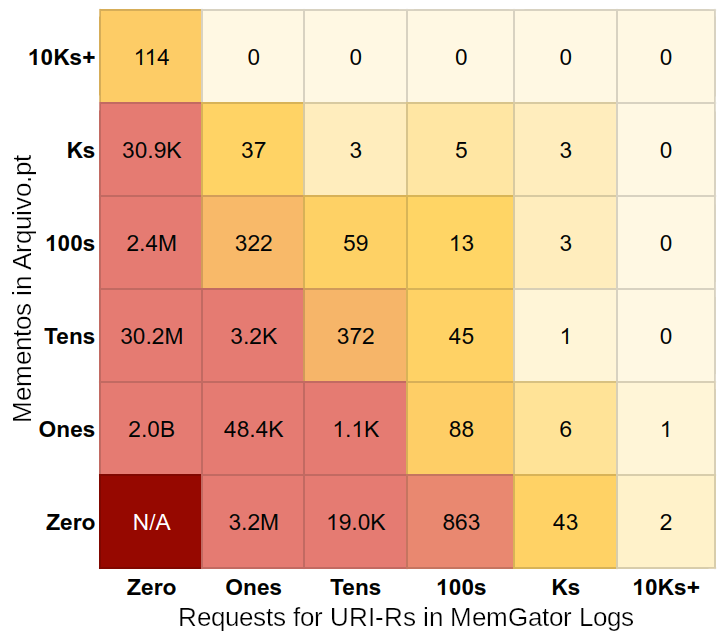}
  \caption{Overlap between archived and accessed resources in \emph{Arquivo.pt}.
           \textnormal{\emph{Ones} denote single digit non-zero numbers (i.e., 1--9), \emph{Tens} denote two digit numbers (i.e., 10--99), and so on.
                       The \emph{Zero} column shows the number of mementos of various \emph{URI-Rs} that are never accessed using \emph{MemGator}.
                       The \emph{Zero} row shows the number of access requests for various \emph{URI-Rs} using \emph{MemGator} that are not archived.
                       The \emph{(Zero, Zero)} cell denotes \emph{N/A} because the number of resources that are neither archived nor accessed is unknown.}
          }
  \label{img:pt-urim-mg-log-frequency}
\end{figure}

Figure~\ref{img:pt-urim-mg-log-frequency} shows a breakdown of what people are looking for in archives and what web archives hold.
The 1.1K entry in the ``Ones'' row and ``Tens'' column shows that there are over a thousand \emph{URI-Rs} that were requested 10--99 times in \emph{MemGator} and each has 1--9 mementos in \emph{Arquivo.pt}.
Large numbers in the ``Zero'' column show there are a lot of mementos that are never requested from \emph{MemGator}.
Similarly, the ``Zero'' row shows there are a lot of requests that have zero mementos in \emph{Arquivo.pt}.
Another way to look at it is that a content-based archive profile will not know about the ``Zero'' row and a usage-based profile will miss out the content in the ``Zero'' column.
Active archives may want to profile their access logs periodically to identify potential seed URIs of frequently requested missing resources that are within the scope of the archive.
Ideally, we would like more activity along the diagonal that passes from the (Zero, Zero) corner, except the corner itself, which suggests there are undetermined number of \emph{URI-Rs} that were never archived or accessed.\looseness=-1

\begin{table}
  \caption{\emph{URI-M} vs. \emph{URI-R} summary of \emph{Arquivo.pt}.}
  \label{tab:pt-urim-urir-dist}
  \begin{tabular}{l | r}
    \toprule
    \textbf{Attributes}                & \textbf{Values} \\
    \midrule
    Unique URI-Rs                      &   1,999,790,376 \\
    Total number of mementos           &   4,923,080,506 \\
    Maximum mementos for any URI-R     &       2,308,634 \\
    Median (and Minimum)               &               1 \\
    Mean mementos per URI-R ($\gamma$) &            2.46 \\
    Standard Deviation                 &           57.20 \\
    Gini Coefficient                   &            0.42 \\
    Pareto Break Point                 &           70/30 \\
    \bottomrule
  \end{tabular}
\end{table}

\begin{figure*}
  \centering
  \subfigure[Percentage of \emph{URI-Rs} by popularity vs. cumulative percentage of mementos.]{
    \label{img:pt-urir-urim-growth}
    \includegraphics[width=0.45\linewidth]{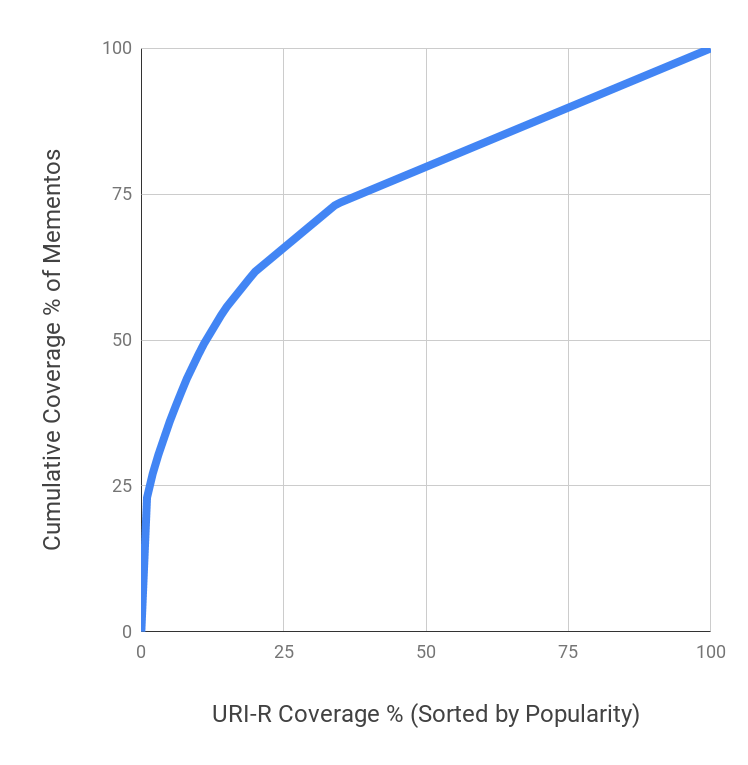}}
  \subfigure[Gini coefficient of memento over \emph{URI-R} population.]{
    \label{img:pt-urir-urim-gini}
    \includegraphics[width=0.45\linewidth]{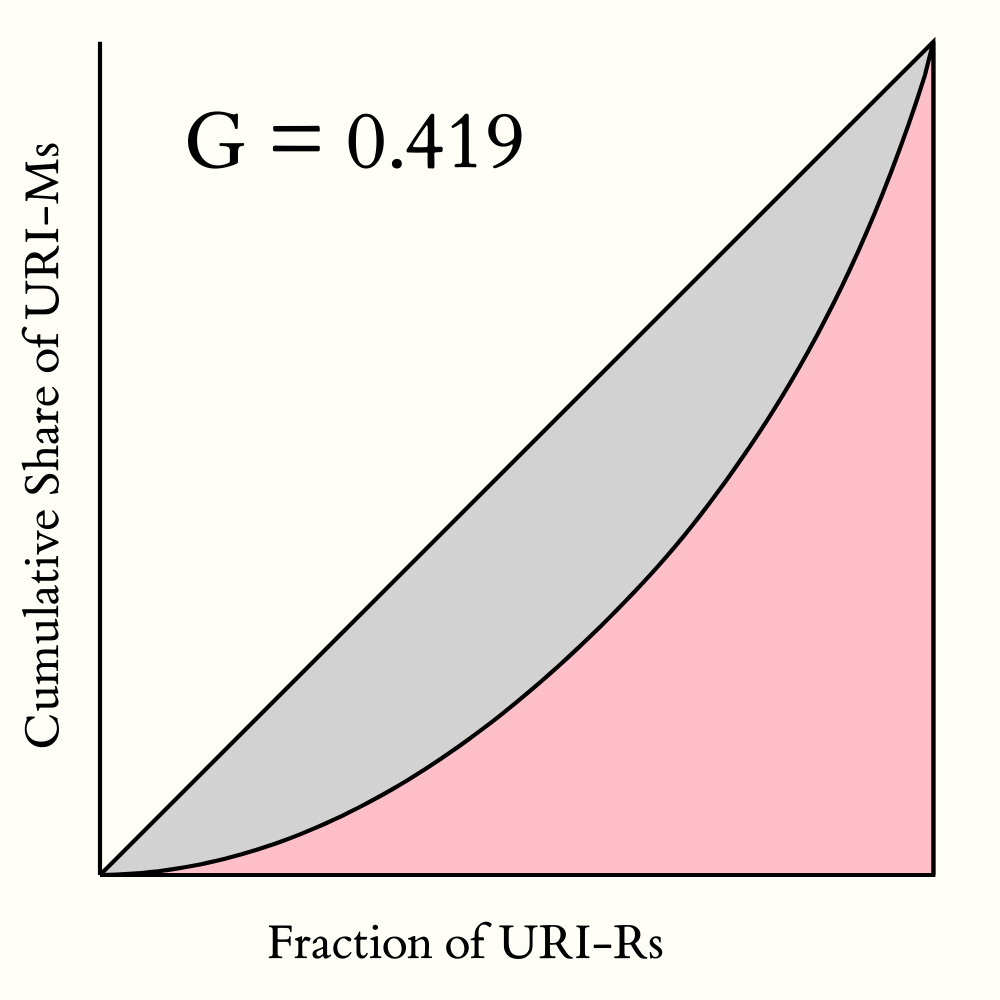}}
  \caption{Distribution of mementos over \emph{URI-Rs} in \emph{Arquivo.pt}.}
  \label{img:urim-urir-distribution}
\end{figure*}

\begin{table}
  \caption{Most archived \emph{URI-Rs} in \emph{Arquivo.pt}.
           \textnormal{Most of these resources are either single pixel blank images or corner graphics used for styling in the pre-CSS3 era.}
  }
  \label{tab:pt-top-urirs}
  \begin{tabular}{l | r}
    \toprule
    \textbf{URIs}                                   & \textbf{URI-Ms} \\
    \midrule
    com,wunderground,icons)/graphics/blank.gif      &       2,308,634 \\
    com,wunderground,icons)/graphics/wuicorner.gif  &         768,250 \\
    pt,ipleiria,inscricoes)/logon.aspx              &         238,292 \\
    com,wunderground,icons)/graphics/wuicorner2.gif &         207,448 \\
    com,lygo)/ly/i/inv/dot\_clear.gif               &         115,221 \\
    com,listbot)/subscribe\_button.gif              &         108,530 \\
    \midrule
    com,wunderground,icons)/* (including top URI-R) &       3,336,086 \\
    com,wunderground,* (41 sub-domains)             &       3,392,676 \\
    \bottomrule
  \end{tabular}
\end{table}

\subsection{Holdings of Arquivo.pt}

Table~\ref{tab:pt-urim-urir-dist} and Figure~\ref{img:urim-urir-distribution} summarize the distribution of \emph{URI-Ms} over \emph{URI-Rs} in \emph{Arquivo.pt}.
Almost 2M unique \emph{URI-Rs} in \emph{Arquivo.pt} have an average of 2.46 mementos per \emph{URI-R} ($\gamma$ value~\cite{alam-ijdl16-profiling}), but this distribution is not uniform.
The top 30\% \emph{URI-Rs} account for 70\% of the mementos, for a \emph{Gini Coefficient} of 0.42~\cite{gini}.
Additionally, the \emph{Median} is one, which means at least half of the \emph{URI-Rs} have only one memento.
Furthermore, the most frequently archived \emph{URI-R} has 2.3M mementos (i.e., 0.05\% of total), so we decided to investigate it further.
Table~\ref{tab:pt-top-urirs} lists the six most archived \emph{URI-Rs}, and they are mostly one pixel clear images and corner graphics primarily used in web designing in the pre-CSS3 era.
The only HTML page that shows up in the top list is a login page.
We further investigated all the mementos from all the subdomains of the top \emph{URI-R}'s domain and found that the \texttt{blank.gif} image was archived out of proportion.
This shows another use for archive profiling -- identifying such unintentional biases due to misconfigured crawling policies or bugs in crawlers' frontier queue management.

\begin{table*}
  \caption{Yearly distribution of \emph{URI-Rs}, \emph{URI-Ms}, and status codes in \emph{Arquivo.pt}.
           \textnormal{The symbol $\gamma$ denotes the ratio of \emph{URI-Ms} vs. \emph{URI-Rs}.
                       Column names with a ``$^+$'' superscript denote cumulative values as yearly data is processed incrementally.
                       While \emph{URI-M$^+$} represents a running total, \emph{URI-R$^+$} does not, because some \emph{URI-Rs} are already seen in previous years.
                       Status codes for the last two years (still in embargo period) do not add up to 100\% because a significant portion of their entries are either \emph{revisit} records or screenshots.}
          }
  \label{tab:yearly-dist}
  \begin{tabular}{l | r r r r r r r r r r r r}
    \toprule
    \textbf{Year} & \textbf{URI-R} & \textbf{URI-R$^+$} & \textbf{URI-M} & \textbf{URI-M$^+$} & \textbf{Dup. URI-R\%} & \textbf{$\gamma$} & \textbf{$\gamma^+$} & \textbf{2xx\%} & \textbf{3xx\%} & \textbf{4xx\%} & \textbf{5xx\%} \\
    \midrule
    1992          &              1 &                  1 &             1  &                  1 &                  0.00 &              1.00 &                1.00 &         100.00 &           0.00 &           0.00 &           0.00 \\
    1993          &              1 &                  2 &             1  &                  2 &                  0.00 &              1.00 &                1.00 &         100.00 &           0.00 &           0.00 &           0.00 \\
    1994          &            128 &                130 &           225  &                227 &                  0.00 &              1.76 &                1.75 &         100.00 &           0.00 &           0.00 &           0.00 \\
    1995          &            642 &                772 &           742  &                969 &                  0.00 &              1.16 &                1.26 &         100.00 &           0.00 &           0.00 &           0.00 \\
    1996          &        110,531 &            111,303 &       126,600  &            127,569 &                  0.00 &              1.15 &                1.15 &          99.96 &           0.01 &           0.00 &           0.00 \\
    1997          &        466,515 &            563,734 &       847,783  &            975,352 &                  3.02 &              1.82 &                1.73 &         100.00 &           0.00 &           0.00 &           0.00 \\
    1998          &        447,042 &            928,112 &       747,114  &          1,722,466 &                 18.49 &              1.67 &                1.86 &          99.23 &           0.77 &           0.00 &           0.00 \\
    1999          &        732,866 &          1,513,381 &     1,233,994  &          2,956,460 &                 20.14 &              1.68 &                1.95 &          76.52 &          10.61 &          12.84 &           0.00 \\
    2000          &      1,710,099 &          2,874,152 &    13,413,518  &         16,369,978 &                 20.43 &              7.84 &                5.70 &          86.99 &           7.24 &           5.73 &           0.00 \\
    2001          &      4,837,012 &          7,286,174 &     7,873,642  &         24,243,620 &                  8.79 &              1.63 &                3.33 &          93.87 &           4.87 &           1.25 &           0.01 \\
    2002          &      7,675,876 &         13,364,488 &    13,048,749  &         37,292,369 &                 20.81 &              1.70 &                2.79 &          90.96 &           5.11 &           3.92 &           0.01 \\
    2003          &     11,043,675 &         21,565,730 &    19,989,725  &         57,282,094 &                 25.74 &              1.81 &                2.66 &          92.12 &           4.45 &           3.41 &           0.03 \\
    2004          &     11,550,512 &         29,460,627 &    22,810,763  &         80,092,857 &                 31.65 &              1.97 &                2.72 &          92.00 &           5.11 &           2.88 &           0.01 \\
    2005          &      9,057,866 &         35,249,604 &    19,839,405  &         99,932,262 &                 36.09 &              2.19 &                2.83 &          93.99 &           3.94 &           2.07 &           0.01 \\
    2006          &      5,979,310 &         39,609,628 &    15,388,836  &        115,321,098 &                 27.08 &              2.57 &                2.91 &          92.33 &           6.29 &           1.37 &           0.01 \\
    2007          &     26,841,427 &         63,396,199 &    43,021,527  &        158,342,625 &                 11.38 &              1.60 &                2.50 &          83.03 &          14.88 &           2.08 &           0.01 \\
    2008          &    113,915,969 &        166,926,098 &   174,996,303  &        333,338,928 &                  9.12 &              1.54 &                2.00 &          85.87 &           8.95 &           6.18 &           0.37 \\
    2009          &    249,069,391 &        383,960,128 &   355,833,394  &        689,172,322 &                 12.86 &              1.43 &                1.79 &          87.37 &           6.55 &           6.49 &           0.36 \\
    2010          &    174,786,328 &        487,044,797 &   352,019,433  &      1,041,191,755 &                 41.02 &              2.01 &                2.14 &          87.39 &           6.83 &           6.49 &           0.42 \\
    2011          &    206,966,813 &        634,061,322 &   465,274,765  &      1,506,466,520 &                 28.97 &              2.25 &                2.38 &          89.13 &           6.21 &           6.99 &           0.58 \\
    2012          &    118,916,669 &        703,235,309 &   200,042,923  &      1,706,509,443 &                 41.83 &              1.68 &                2.43 &          87.79 &           6.66 &           7.96 &           0.46 \\
    2013          &    174,913,693 &        827,924,633 &   236,583,969  &      1,943,093,412 &                 28.71 &              1.35 &                2.35 &          84.03 &           7.28 &          10.90 &           0.57 \\
    2014          &    430,555,712 &      1,166,054,663 &   536,560,181  &      2,479,653,593 &                 21.47 &              1.25 &                2.13 &          80.50 &           7.10 &          13.47 &           0.52 \\
    2015          &    558,504,002 &      1,563,688,006 & 1,087,680,516  &      3,567,334,109 &                 28.80 &              1.95 &                2.28 &          78.32 &           5.12 &          17.75 &           0.32 \\
    2016          &    719,889,903 &      1,999,522,571 & 1,353,786,928  &      4,921,121,037 &                 39.46 &              1.88 &                2.46 &          73.20 &           6.46 &          20.78 &           1.30 \\
    2017          &        685,097 &      1,999,687,103 &     1,111,999  &      4,922,233,036 &                 75.98 &              1.62 &                2.46 &          57.82 &           5.44 &           7.89 &           0.22 \\
    2018          &        106,186 &      1,999,790,376 &       847,470  &      4,923,080,506 &                  2.74 &              7.98 &                2.46 &          22.07 &           5.63 &           1.38 &           0.00 \\
    \midrule
    \textbf{All}  &  1,999,790,376 &      1,999,790,376 & 4,923,080,506  &      4,923,080,506 &                  0.00 &              2.46 &                2.46 &          80.74 &           6.42 &          13.86 &           0.66 \\
    \bottomrule
  \end{tabular}
\end{table*}

\begin{figure}
  \includegraphics[width=\linewidth]{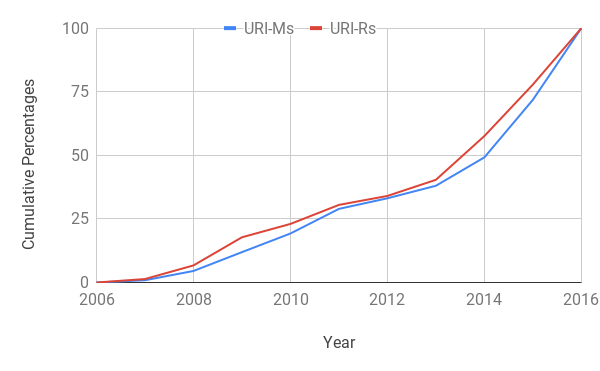}
  \caption{Cumulative growth of \emph{URI-Rs} and \emph{URI-Ms} in \emph{Arquivo.pt}.
           \textnormal{Almost half of the mementos are captured in the last two active years alone.}
          }
  \label{img:pt-active-urir-urim-growth-curve}
\end{figure}

Furthermore, we partitioned \emph{Arquivo.pt}'s index into yearly buckets for analysis as shown in Table~\ref{tab:yearly-dist}.
Data prior to year 2008 is mostly donated from other sources in the form of many small files, as \emph{Arquivo.pt} was not yet established.
However, when everything is put together it looks like the archiving activity took off significantly in 2007.
Low numbers in years 2017 and 2018 are due to \emph{Arquivo.pt}'s embargo policy.
It shows that \emph{Arquivo.pt}'s collection is growing with a healthy pace by mostly collecting new \emph{URI-Rs} as well as revisiting on an average 26\% of older ones on a yearly basis.
We expected $\gamma$ would change gradually over time, but years 2000 and 2018 had significantly high values with respect to other years.
So, we looked for the possibility of increased \texttt{3xx} status codes in those years as a potential source of increase in $\gamma$ (e.g., \texttt{http} \emph{URIs} redirecting to corresponding \texttt{https} version), but we did not see any correlation there.
However, the data for these years seems to have come from another source and overall they are insignificant, hence, the cumulative $\gamma^+$ is fairly stable between 2 and 3.
We noted a significant and steady growth in \texttt{4xx} status codes which has crossed the 20\% mark in year 2016.
Status codes for the last two years (still in embargo period) do not sum up to 100\% because a significant portion of their entries are either \emph{revisit} records or screenshots that do not report status codes.
In Figure~\ref{img:pt-active-urir-urim-growth-curve} we plotted a cumulative growth graph of both \emph{URI-Ms} and \emph{URI-Rs} to see the shape~\cite{shapeait} of \emph{Arquivo.pt} during the active region.
Their archiving rate is increasing over time as almost half of the total mementos were archived in the last two active years alone.

\begin{table}
  \caption{Unique items with exact \emph{Host} and \emph{Path} depths.}
  \label{tab:host-path-depth-exact}
  \begin{tabular}{l | r r r}
    \toprule
    \textbf{Depth} & \textbf{Host (Domains)} & \textbf{Host (HxPx)} & \textbf{Path (HxPx)} \\
    \midrule
    0              &                       1 &                    1 &            4,456,831 \\
    1              &                     119 &                6,479 &          113,022,403 \\
    2              &               1,949,845 &          508,607,506 &          225,489,773 \\
    3              &               2,097,254 &          429,000,297 &          334,455,187 \\
    4              &               1,316,005 &          161,912,251 &          174,429,887 \\
    5              &                 234,110 &           21,825,084 &          127,484,179 \\
    6              &                  95,492 &            7,935,125 &           68,578,693 \\
    7              &                  28,121 &            3,252,943 &           45,819,300 \\
    8              &                  64,716 &            3,722,893 &           22,178,800 \\
    9              &                  55,801 &            2,660,529 &           15,553,102 \\
    10             &                       5 &                   50 &            6,596,158 \\
    11+            &                       5 &                   12 &              858,856 \\
    \midrule
    \textbf{Total} &               5,841,473 &        1,138,923,169 &        1,138,923,169 \\
    \bottomrule
  \end{tabular}
\end{table}

\subsection{The Shape of Archived URI Tree}

To understand the shape of the \emph{URI Keys} tree in \emph{MementoMap} we first investigated the number of unique \emph{Domains} and \emph{HxPx Keys} that have certain host or path depths as shown in Table~\ref{tab:host-path-depth-exact}.
These numbers are relative to the size of the \emph{Arquivo.pt} index, but we believe a similar trend should be seen in other archives, unless their collection is manually curated and crawled using a more or less capable tool than what is currently being used by many large web archives~\cite{heritrix}.
There were some outliers in the data that showed a host depth of up to 15 and path depths up to 130, but those were very few in number.
These numbers gave us a good starting point to decide how deep we need to analyze hosts and paths for profiling.

\begin{figure*}
  \centering
  \subfigure[\emph{Parents} and \emph{Children} at each \emph{Host} depth.
             All the terminating host nodes at each level lead to the root path (i.e., \emph{P0}) shown at the bottom.]{
    \label{img:tree-shape-host}
    \includegraphics[width=0.48\linewidth]{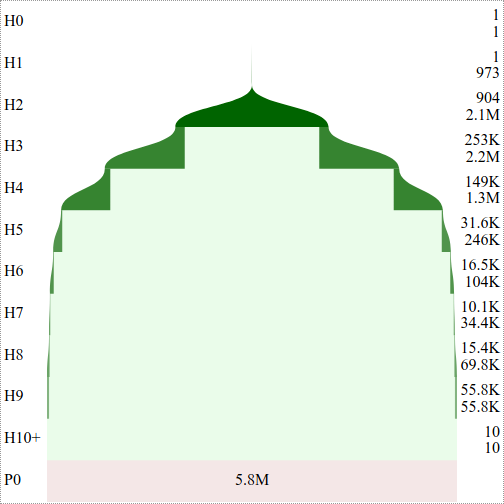}}
  \subfigure[\emph{Parents} and \emph{Children} at each \emph{Path} depth.
             The root path (i.e., \emph{P0}) shown at the top is scaled 70 times down as compared with the bottom row of the \emph{Host} segment tree.]{
    \label{img:tree-shape-path}
    \includegraphics[width=0.48\linewidth]{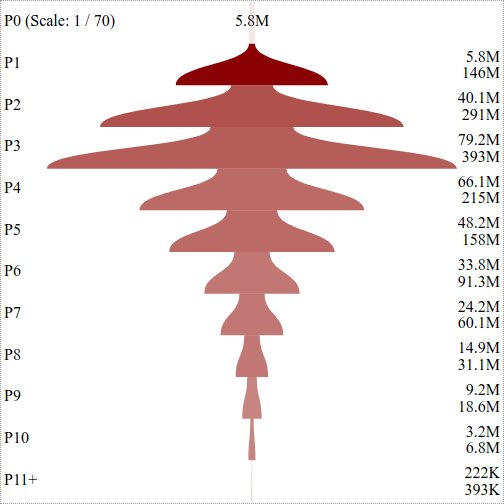}}
  \caption{The shape of \emph{HxPx Key} tree of \emph{Arquivo.pt}.
           \textnormal{Labels on the left denote \emph{Host} and \emph{Path} depths.
                       Corresponding pair of labels on the right denote number of \emph{Parents} and \emph{Children} respectively.
                       Darker nodes have higher number of \emph{Mean Children}.
                       \emph{Host} and \emph{Path} segments are plotted separately with different scales while the bottom row of the \emph{Host} segment corresponds to the top row of the \emph{Path} segment.}
          }
  \label{img:tree-shape}
\end{figure*}

Figure~\ref{img:tree-shape} shows the shape of the total 1,138,923,169 unique \emph{HxPx Keys} of \emph{Arquivo.pt}'s current index put together in the form of a tree as the \emph{URI Key} space changes on each host and path depth.
The tree is broken down in host and path segments (i.e., Figure~\ref{img:tree-shape-host} and \ref{img:tree-shape-path}) instead of one continuous tree and the latter is scaled down 70 times as compared with the host segment to ensure that the shape of path segment is distinguishable from one depth to the next.
In the host segment, at each host level (after \emph{H1}) a significant portion leads to \emph{P0} (i.e., root path), but the remainder has further children host segments (i.e., sub-domains).
Figure~\ref{img:tree-shape-host} shows that hostnames with depth more than four (i.e., \emph{H5} and beyond) are significantly small in number.
In the path segment, at each level a significant portion terminates, but the remainder branches out into deeper path segments.
The shape of the path segment in Figure~\ref{img:tree-shape-path} shows that the tree starts to shrink from \emph{P4} and the bulk tree is around \emph{P3}.
Any effort to reduce the \emph{URI Key} space near this level can significantly reduce the \emph{Relative Cost}.

\begin{table*}
  \caption{\emph{Host} and \emph{Path} depth statistics of unique \emph{HxPx Keys} in \emph{Arquivo.pt}.
           \textnormal{Sorted \emph{HxPx Keys} no shorter than a given depth are chopped at that depth, number of occurrences of these keys is \emph{Count}, their total is \emph{Sum}, and various other statistical measures are reported based on these numbers.
                       The \emph{RedQ} value is calculated using Equation~\ref{eq:redq}, \emph{Parents} is the number of non-terminal nodes of the previous depth, \emph{Children} is the number of unique nodes at a given depth, and \emph{MeanChld} is the average number of \emph{Children} per \emph{Parent}.}
          }
  \label{tab:host-path-depth-stats}
  \begin{tabular}{l | r r r r r r r r r r}
    \toprule
    \textbf{Depth} & \textbf{Count} &  \textbf{Sum} & \textbf{Max} & \textbf{Mean} & \textbf{Med.} & \textbf{StdDev} & \textbf{RedQ} & \textbf{Parents} & \textbf{Children} & \textbf{MeanChld} \\
    \midrule
    H1             &            973 & 1,138,923,169 &  616,372,626 &  1,170,527.41 &           930 &   21,620,107.00 &       1.00000 &                1 &               973 &            973.00 \\
    H2             &      2,068,333 & 1,138,916,690 &  109,176,956 &        550.64 &             5 &       91,308.66 &       0.99818 &              904 &         2,068,333 &          2,287.98 \\
    H3             &      2,158,880 &   630,309,184 &   51,849,377 &        291.96 &             7 &       37,641.59 &       0.55153 &          253,091 &         2,158,880 &              8.53 \\
    H4             &      1,329,137 &   201,308,887 &    3,765,122 &        151.46 &            10 &        4,797.10 &       0.17559 &          148,589 &         1,329,137 &              8.95 \\
    H5             &        245,881 &    39,396,636 &      376,969 &        160.23 &             5 &        3,420.96 &       0.03438 &           31,635 &           245,881 &              7.77 \\
    H6             &        103,579 &    17,571,552 &      105,591 &        169.64 &            27 &        1,106.03 &       0.01534 &           16,496 &           103,579 &              6.28 \\
    H7             &         34,380 &     9,636,427 &       19,572 &        280.29 &            20 &          450.16 &       0.00843 &           10,061 &            34,380 &              3.42 \\
    H8             &         69,829 &     6,383,484 &          535 &         91.42 &           120 &           45.75 &       0.00554 &           15,359 &            69,829 &              4.55 \\
    H9             &         55,811 &     2,660,591 &           80 &         47.67 &            56 &            19.6 &       0.00229 &           55,811 &            55,811 &              1.00 \\
    H10+           &             10 &            62 &           19 &          6.20 &             2 &            6.51 &       0.00000 &               10 &                10 &              1.00 \\
    P0             &      5,841,503 & 1,138,923,169 &    2,264,623 &        194.97 &             7 &        3,059.43 &       0.99487 &        5,841,503 &         5,841,503 &              1.00 \\
    P1             &    145,687,459 & 1,134,466,338 &    2,242,344 &          7.79 &             1 &          376.64 &       0.86817 &        5,828,059 &       145,687,459 &             25.00 \\
    P2             &    290,761,965 & 1,021,443,935 &      603,840 &          3.51 &             1 &          130.76 &       0.64156 &       40,130,355 &       290,761,965 &              7.25 \\
    P3             &    392,635,328 &   795,954,162 &      565,043 &          2.03 &             1 &           78.14 &       0.35412 &       79,234,027 &       392,635,328 &              4.96 \\
    P4             &    215,251,988 &   461,498,975 &      512,098 &          2.14 &             1 &           80.01 &       0.21621 &       66,059,544 &       215,251,988 &              3.26 \\
    P5             &    158,256,277 &   287,069,088 &      512,098 &          1.81 &             1 &           65.72 &       0.11310 &       48,163,114 &       158,256,277 &              3.29 \\
    P6             &     91,334,214 &   159,584,909 &       50,384 &          1.75 &             1 &            22.3 &       0.05993 &       33,776,599 &        91,334,214 &              2.70 \\
    P7             &     60,099,825 &    91,006,216 &       44,114 &          1.51 &             1 &           17.24 &       0.02714 &       24,201,781 &        60,099,825 &              2.48 \\
    P8             &     31,101,768 &    45,186,916 &       24,631 &          1.45 &             1 &           15.54 &       0.01237 &       14,890,308 &        31,101,768 &              2.09 \\
    P9             &     18,601,197 &    23,008,116 &       10,247 &          1.24 &             1 &            9.74 &       0.00387 &        9,233,634 &        18,601,197 &              2.01 \\
    P10            &      6,817,122 &     7,455,014 &        5,858 &          1.09 &             1 &            9.36 &       0.00056 &        3,206,260 &         6,817,122 &              2.13 \\
    P11+           &        858,772 &       858,856 &            2 &          1.00 &             1 &            0.01 &       0.00000 &          222,432 &           392,565 &              1.76 \\
    \bottomrule
  \end{tabular}
\end{table*}

\begin{figure*}
  \centering
  \subfigure[Global \emph{HxPx} reduction rate at \emph{Host}.]{
    \label{img:host-reduction-rate}
    \includegraphics[width=0.47\linewidth]{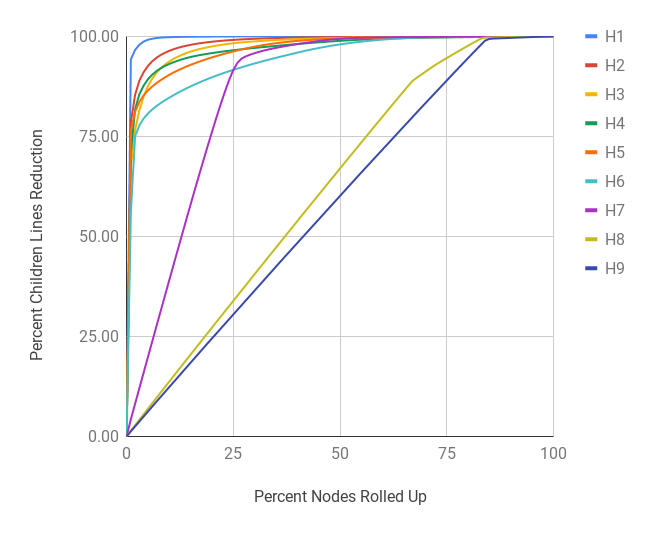}}
  \subfigure[Global \emph{HxPx} reduction rate at \emph{Path}.]{
    \label{img:path-reduction-rate}
    \includegraphics[width=0.47\linewidth]{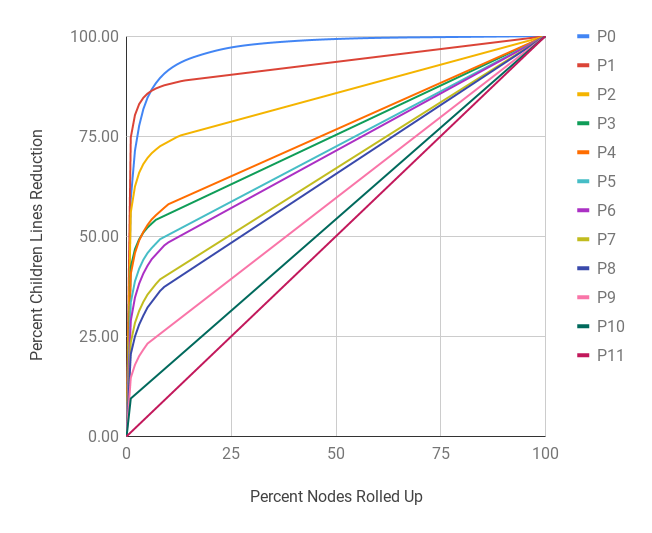}}
  \subfigure[Incremental \emph{Host} children reduction.]{
    \label{img:host-reduction-rate-incr}
    \includegraphics[width=0.47\linewidth]{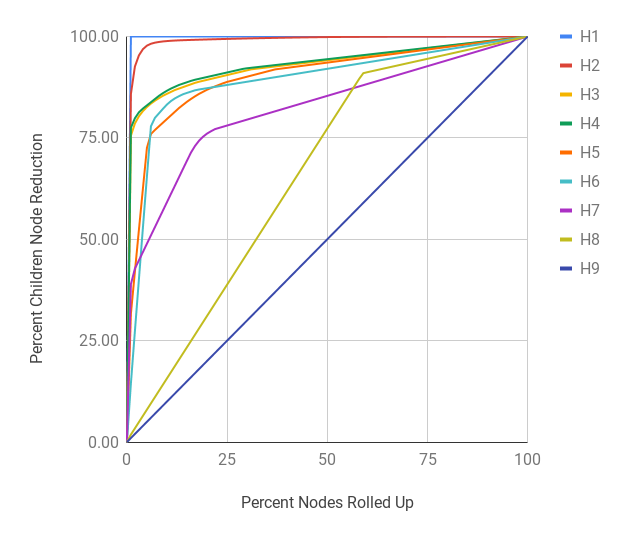}}
  \subfigure[Incremental \emph{Path} children reduction.]{
    \label{img:path-reduction-rate-incr}
    \includegraphics[width=0.47\linewidth]{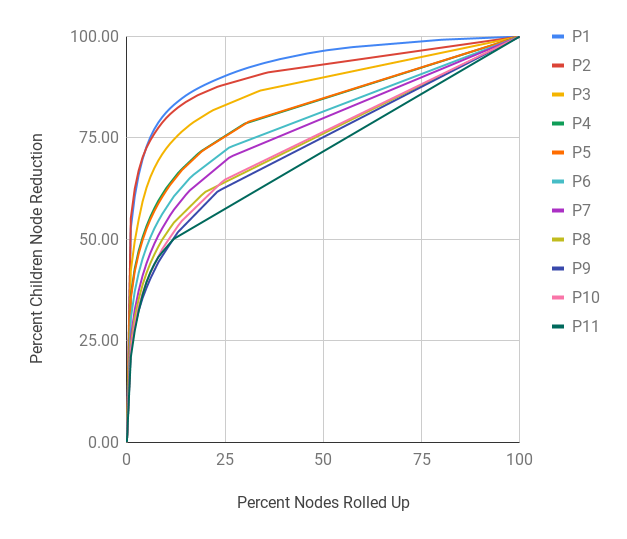}}
  \caption{Global and incremental \emph{Host} and \emph{Path} segment reduction.
           \textnormal{Global reductions describe the change in the total number of \emph{HxPx Keys} (or the size of sub-trees) when keys are rolled up at a given \emph{Host} or \emph{Path} depth.
                       Incremental children reductions describe the change caused by roll ups of immediate children nodes into their corresponding parent nodes at a given \emph{Host} or \emph{Path} depth.
                       Nodes with larger sub-trees and children counts in the two cases respectively are rolled up first.}
          }
  \label{img:reduction-rates}
\end{figure*}

Table~\ref{tab:host-path-depth-stats} is based on the total 1,138,923,169 unique \emph{HxPx Keys} of \emph{Arquivo.pt}'s current index.
For example, the \emph{H3} (see Figure~\ref{img:surt-tree} for naming convention) row means there are a total of 2,158,880 unique \emph{H3} prefixes that cover a sum of 630,309,184 \emph{HxPx Keys} of which the most popular prefix covers 51,849,377 keys alone.
The \emph{Mean} number of keys per prefix at \emph{H3} is 291.96 with a \emph{Median} of 7 and \emph{Standard Deviation} of 37,641.59.
The \emph{RedQ (Reduction Coefficient)} column represents a derived quantity that we defined as the amount of reduction in keys it would cause if \emph{HxPx Keys} longer than a given depth are stripped off at that depth and only counted reduced unique prefixes.
This can be calculated using Equation~\ref{eq:redq} at depth $d$ where $|\text{HxPx Keys}_{\geq d}|$ is the number of \emph{HxPx Keys} with depth $\geq d$ and $|\text{URI Keys}_d|$ is the number of unique partial \emph{URI Keys} stripped at depth $d$ (reported under the \emph{Sum} and \emph{Count} columns of Table~\ref{tab:host-path-depth-stats} respectively).
Figures~\ref{img:host-reduction-rate} and \ref{img:path-reduction-rate} show the cumulative reduction as the top most frequent keys are rolled up at a host and path depth respectively.
Furthermore, there are 253,091 nodes in the tree one depth above (i.e., \emph{H2}) that lead to 2,158,880 nodes at the current depth.
While the \emph{Mean Child} count at \emph{H3} is 8.53, the distribution is not uniform.
Figures~\ref{img:host-reduction-rate-incr} and \ref{img:path-reduction-rate-incr} show the cumulative reduction in immediate children count as the most popular parents leading to the current depth are rolled up incrementally from bottom up.
The purpose of the \emph{Reduction Coefficient} is to understand the impact and importance of various host and path depths globally while the \emph{Mean Child} count gives an estimate of a more localized impact at a given depth.
For this work we have used the latter as a factor to decide when to roll a sub-tree up while compacting a \emph{MementoMap}.
Rolling the sub-tree up at \emph{H1}, \emph{H2}, and \emph{P0} are not applicable for evaluation here because \emph{H1} means shrinking everything into a single record of ``*'' key, \emph{H2} would require out-of-band information (because not every \emph{TLD} is equally popular), and \emph{P0} being the root of the path has nothing to roll up into (though compaction might happen in the relevant host segment independently).
We fit the remaining values of \emph{Mean Child} count on Power Law~\cite{powerlaw} curves (other curve fittings are also possible) for both host and path segments to find $a$ and $k$ parameters and use these empirical values for compaction decision making.\looseness=-1

\begin{equation}
  \label{eq:redq}
  \text{RedQ}_d = \frac{|\text{HxPx Keys}_{\geq d}| - |\text{URI Keys}_d|}{|\text{HxPx Keys}|}
\end{equation}

\begin{table*}
  \caption{\emph{MementoMap} generation, compaction, and lookup statistics for \emph{Arquivo.pt}.
           \textnormal{Output of one step is used as the input of the next step in a chain as the next step has at least one smaller weight.
                       The first record was created using some \emph{Linux} commands instead of the script, that is why some values are reported as \emph{N/A}.}
          }
  \label{tab:mmap-compact}
  \begin{tabular}{l r r | r r r r r r r}
    \toprule
    \textbf{Input}         & \textbf{$W_h$} & \textbf{$W_p$} & \textbf{Lines} & \textbf{Size (bytes)} & \textbf{Gzipped (MB)} & \textbf{Rollups} & \textbf{Time (sec)} & \textbf{RelCost} & \textbf{Accuracy} \\
    \midrule
    CDXJ                   &       $\infty$ &       $\infty$ &    447,107,301 &        30,753,644,382 &                 3,449 &              N/A &                 N/A &            0.464 &             0.946 \\
    $H_{\infty}P_{\infty}$ &           4.00 &           4.00 &     27,010,037 &         1,443,292,676 &                   218 &        4,574,305 &               8,643 &            0.028 &             0.646 \\
    $H_{4.00}P_{4.00}$     &           4.00 &           2.00 &     14,143,676 &           662,171,623 &                   119 &          703,394 &                 507 &            0.015 &             0.600 \\
    $H_{4.00}P_{2.00}$     &           4.00 &           1.00 &      7,528,548 &           315,946,553 &                    63 &          537,341 &                 264 &            0.008 &             0.539 \\
    $H_{4.00}P_{1.00}$     &           4.00 &           0.50 &      4,269,344 &           162,132,599 &                    35 &          483,779 &                 151 &            0.004 &             0.482 \\
    $H_{4.00}P_{0.50}$     &           4.00 &           0.25 &      3,054,686 &           107,784,353 &                    24 &          411,843 &                  87 &            0.003 &             0.426 \\
    $H_{4.00}P_{0.25}$     &           4.00 &           0.00 &      1,673,784 &            40,446,417 &                    11 &        1,411,579 &                  70 &            0.002 &             0.275 \\
    $H_{4.00}P_{4.00}$     &           2.00 &           4.00 &     24,937,984 &         1,316,371,599 &                   205 &            9,572 &                 500 &            0.026 &             0.626 \\
    $H_{2.00}P_{4.00}$     &           2.00 &           2.00 &     12,867,647 &           585,142,758 &                   111 &          669,670 &                 468 &            0.013 &             0.588 \\
    $H_{2.00}P_{2.00}$     &           2.00 &           1.00 &      6,584,376 &           257,905,766 &                    58 &          512,413 &                 241 &            0.007 &             0.525 \\
    $H_{2.00}P_{1.00}$     &           2.00 &           0.50 &      3,615,997 &           121,452,813 &                    32 &          458,681 &                 124 &            0.004 &             0.472 \\
    $H_{2.00}P_{0.50}$     &           2.00 &           0.25 &      2,542,869 &            76,274,453 &                    21 &          349,700 &                  70 &            0.003 &             0.422 \\
    $H_{2.00}P_{0.25}$     &           2.00 &           0.00 &      1,529,328 &            33,658,544 &                    10 &        1,171,377 &                  56 &            0.002 &             0.270 \\
    $H_{2.00}P_{4.00}$     &           1.00 &           4.00 &     23,840,710 &         1,252,548,065 &                   196 &            4,671 &                 466 &            0.025 &             0.581 \\
    $H_{1.00}P_{4.00}$     &           1.00 &           2.00 &     12,313,036 &           555,628,348 &                   107 &          640,163 &                 448 &            0.013 &             0.549 \\
    $H_{1.00}P_{2.00}$     &           1.00 &           1.00 &      6,307,180 &           244,402,690 &                    56 &          489,942 &                 232 &            0.007 &             0.501 \\
    $H_{1.00}P_{1.00}$     &           1.00 &           0.50 &      3,465,689 &           114,755,789 &                    30 &          439,647 &                 116 &            0.004 &             0.453 \\
    $H_{1.00}P_{0.50}$     &           1.00 &           0.25 &      2,437,451 &            71,797,863 &                    20 &          333,087 &                  67 &            0.003 &             0.403 \\
    $H_{1.00}P_{0.25}$     &           1.00 &           0.00 &      1,474,541 &            31,881,496 &                    10 &        1,117,830 &                  58 &            0.002 &             0.261 \\
    $H_{1.00}P_{4.00}$     &           0.50 &           4.00 &     22,315,969 &         1,162,107,385 &                   184 &            6,516 &                 447 &            0.023 &             0.540 \\
    $H_{0.50}P_{4.00}$     &           0.50 &           2.00 &     11,729,408 &           525,115,243 &                   101 &          594,779 &                 420 &            0.012 &             0.520 \\
    $H_{0.50}P_{2.00}$     &           0.50 &           1.00 &      6,056,959 &           232,945,804 &                    53 &          461,516 &                 218 &            0.006 &             0.476 \\
    $H_{0.50}P_{1.00}$     &           0.50 &           0.50 &      3,342,092 &           109,798,250 &                    29 &          417,912 &                 112 &            0.003 &             0.433 \\
    $H_{0.50}P_{0.50}$     &           0.50 &           0.25 &      2,358,976 &            68,957,985 &                    20 &          316,782 &                  65 &            0.002 &             0.388 \\
    $H_{0.50}P_{0.25}$     &           0.50 &           0.00 &      1,434,084 &            30,800,396 &                     9 &        1,071,071 &                  51 &            0.001 &             0.253 \\
    $H_{0.50}P_{4.00}$     &           0.25 &           4.00 &     21,197,676 &         1,096,034,790 &                   174 &            9,533 &                 416 &            0.022 &             0.511 \\
    $H_{0.25}P_{4.00}$     &           0.25 &           2.00 &     11,217,682 &           498,573,523 &                    97 &          558,528 &                 392 &            0.012 &             0.495 \\
    $H_{0.25}P_{2.00}$     &           0.25 &           1.00 &      5,842,652 &           223,237,207 &                    51 &          435,916 &                 204 &            0.006 &             0.461 \\
    $H_{0.25}P_{1.00}$     &           0.25 &           0.50 &      3,241,589 &           105,791,213 &                    28 &          398,097 &                 109 &            0.003 &             0.420 \\
    $H_{0.25}P_{0.50}$     &           0.25 &           0.25 &      2,298,413 &            66,763,014 &                    19 &          302,762 &                  64 &            0.002 &             0.377 \\
    $H_{0.25}P_{0.25}$     &           0.25 &           0.00 &      1,404,993 &            30,018,340 &                     9 &        1,031,775 &                  53 &            0.001 &             0.249 \\
    $H_{0.25}P_{4.00}$     &           0.00 &           4.00 &     17,391,655 &           882,144,079 &                   142 &          118,082 &                 392 &            0.018 &             0.391 \\
    $H_{0.00}P_{4.00}$     &           0.00 &           2.00 &      9,453,810 &           410,205,661 &                    81 &          560,039 &                 324 &            0.010 &             0.385 \\
    $H_{0.00}P_{2.00}$     &           0.00 &           1.00 &      5,054,662 &           187,327,280 &                    43 &          471,696 &                 179 &            0.005 &             0.373 \\
    $H_{0.00}P_{1.00}$     &           0.00 &           0.50 &      2,901,796 &            91,419,782 &                    25 &          440,818 &                  95 &            0.003 &             0.354 \\
    $H_{0.00}P_{0.50}$     &           0.00 &           0.25 &      2,107,245 &            59,036,815 &                    17 &          366,330 &                  57 &            0.002 &             0.326 \\
    $H_{0.00}P_{0.25}$     &           0.00 &           0.00 &      1,339,475 &            27,946,167 &                     8 &          986,664 &                  48 &            0.001 &             0.236 \\
    \bottomrule
  \end{tabular}
\end{table*}

\subsection{MementoMap Cost and Accuracy}

Web archives are messy collections that contain many malformed records often caused by configuration issues in web servers, poorly written web applications, bugs in archiving tools, incompatible file transformations, or even security vulnerabilities~\cite{methodsupport}.
Archive profiling can uncover some of these as we found many malformed \emph{MIME-Type}\footnote{\url{https://gist.github.com/ibnesayeed/bb167fe19c5719d87c1c1f665001d44b}} and \emph{Status Code}\footnote{\url{https://gist.github.com/ibnesayeed/7307f0bf1783357db99f8b2357249dd0}} entries in \emph{Arquivo.pt}.

To run our experiments we decided to filter only the clean records out from these \emph{CDXJ} files.
We further limited our scope to only \emph{HTML} pages that returned a \texttt{200} status code.
Additionally, we excluded any \texttt{robots.txt} and \texttt{sitemap.xml} files that were served wrongly as ``\texttt{text/html}''.
With these filters in place we reduced mementos by almost half of the total index size to only 2,671,653,766.
Now, there are 962,832,513 filtered unique \emph{URI-Rs}, which means the $\gamma$ value is increased slightly to 2.77.
Also, the \emph{HxPx Keys} count is reduced to 447,107,301, which is 39\% of the overall number.
From these keys we created the baseline \emph{MementoMap} with compressed file size of 3.4G (as shown in the first record of Table~\ref{tab:mmap-compact}) which is already reduced to 1.3\% of the original index size.
This baseline \emph{MementoMap} has 46.4\% \emph{Relative Cost} (i.e., the ratio of reduced number of unique lookup keys vs. number of unique \emph{URI-Rs}) that yields 94.6\% \emph{Accuracy}.

In the next step we supplied this baseline \emph{MementoMap} as input for compaction with host and path compaction weights $W_h=4.00$ and $W_p=4.0$ respectively.
These weights are multiplied by their corresponding estimated \emph{Mean Child} value at each depth to find the cutoff number when the sub-tree is to be rolled up.
A small weight will roll the sub-tree up more aggressively than a large value, resulting in a more compact \emph{MementoMap}.
This process produced a \emph{MementoMap} with only 27,010,037 lines (i.e., 6.0\% of the baseline or 2.8\% \emph{Relative Cost}) after going through 4,574,305 recursive roll ups.
The process took 2.4 hours to complete on our Network File System (NFS) storage.
The time taken to complete the compaction process is a function of the number of lines to process from the input, number of lines to be written out, and the number of roll ups to occur (along with the read and write speeds of the disk).
Since the process is I/O intensive, using faster storage can reduce the time significantly, which we verified by repeating the experiment on \emph{TMPFS}~\cite{tmpfs}.
We generated 36 variations of \emph{MementoMaps} with all possible pairs of $W_h$ and $W_p$ weights from values 4.00, 2.00, 1.00, 0.50, 0.25, and 0.00 as shown in Table~\ref{tab:mmap-compact}.
To generate \emph{MementoMaps} with smaller weights we used \emph{MementoMaps} of immediate larger weight pairs as inputs (e.g., input one with $W_h=2.00,W_p=0.50$ to generate one with $W_h=1.00,W_p=0.25$).
This technique of chaining the output as input to the next step reduced the generation time for subsequent \emph{MementoMaps} from hours to a few minutes and also illustrated that \emph{MementoMaps} can easily be compacted further when needed.

\begin{figure}
  \includegraphics[width=\linewidth]{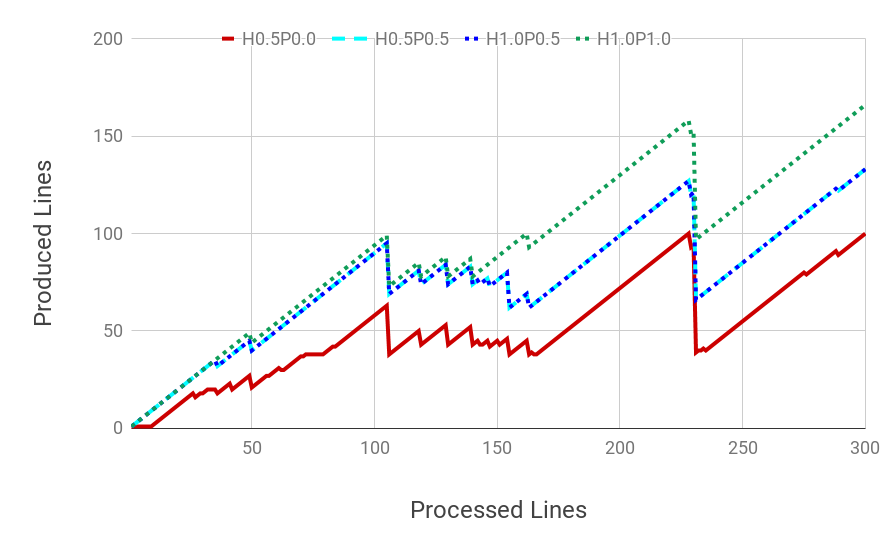}
  \caption{Growth of compacted \emph{MementoMap} vs. lines processed from an input \emph{MementoMap}.
           \textnormal{This plot illustrates a very small portion of the entire process to highlight the compaction behavior at a micro level.
                       The size of the output \emph{MementoMap} decreases each time a roll up happens.
                       A roll up at smaller depth often reduces the size more significantly.}
          }
  \label{img:processed-produced-lines}
\end{figure}

Figure~\ref{img:processed-produced-lines} shows a portion of the roll up activity during the compaction process.
The size of the output grows linearly, but on a micro-scale whenever there is a roll up activity, the output size goes down depending on at what depth roll up happened and how big of a sub-tree was affected.

\begin{figure}
  \includegraphics[width=\linewidth]{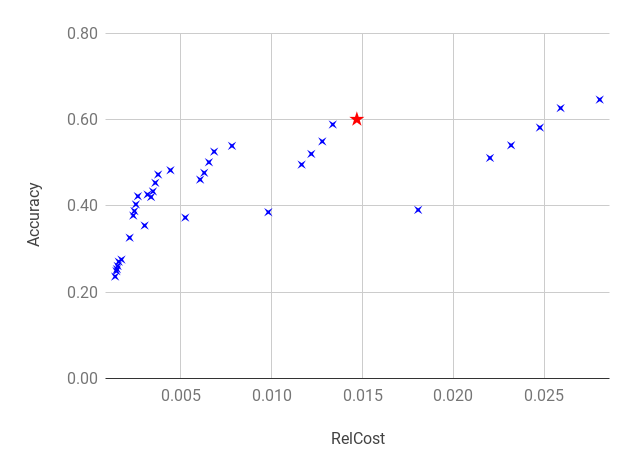}
  \caption{\emph{Relative Cost} vs. \emph{Lookup Routing Accuracy}.
           \textnormal{A \emph{MementoMap} generated/compacted using $W_h=4.00$ and $W_p=2.0$ yielded 60\% \emph{Accuracy} with only 1.5\% \emph{Relative Cost}.}
          }
  \label{img:pt-mg-relcost-accuracy}
\end{figure}

Finally, we used \emph{MemGator} logs to perform lookup in these 36 \emph{MementoMaps} generated with different host and path weight pairs to see how well they perform.
Figure~\ref{img:pt-mg-relcost-accuracy} shows the \emph{Relative Cost} and corresponding \emph{Lookup Routing Accuracy} of these \emph{MementoMaps}.
The \emph{Accuracy} here is defined as the ratio of correctly identified \emph{URIs} for their presence or absence vs. all the lookup \emph{URIs}.
In this experiment \emph{MementoMaps} with weights $W_h=4.00,W_p=2.00$ and $W_h=2.00,W_p=2.00$ yielded about 60\% \emph{Routing Accuracy} with less 1.5\% \emph{Relative Cost} without any false negatives (i.e., 100\% \emph{Recall}).
Since \emph{Arquivo.pt} had only a 3.35\% hit rate in the past three years, \emph{MemGator} could have avoided almost 60\% of the wasted traffic to \emph{Arquivo.pt} without missing any good results if \emph{Arquivo.pt} were to advertise its holdings via a small \emph{MementoMap} of about 111MB in size.
The accuracy can further be improved by 1) exploring other optimal configurations for sub-tree pruning, 2) generating \emph{MementoMaps} with the full index, not just a sample, and 3) including entries for absent resources from the ``Zero'' row of the Figure~\ref{img:pt-urim-mg-log-frequency}.

\section{Conclusions and Future Work}

In this work we proposed \emph{MementoMap}, a flexible and adaptive framework to express holdings of a web archive efficiently.
We described a simple, yet extensible, file format suitable for \emph{MementoMap} and some other use cases.
We extended traditional \emph{SURT} format to support wildcards for partial \emph{URI Keys}.
We analyzed more than three years of \emph{MemGator} logs to understand the response behavior of 14 public web archives.
We used the complete index of 5B mementos in the \emph{Arquivo.pt} as a case study, learned some generalizable behaviors of \emph{URIs} in web archives, described \emph{Arquivo.pt}'s holdings in different ways, and created \emph{MementoMaps} of varying sizes from it for evaluation.
We designed a single-pass, memory-efficient, and parallelization-friendly algorithm to compact a large \emph{MementoMap} into a small one iteratively, based on user-specified parameters to accommodate different needs and available resources.
We also implemented a time-and memory-efficient lookup method using binary search on \emph{MementoMap} files on disk by leveraging the fact that \emph{MementoMaps} are in a lexicographical order.
Finally, we evaluated the effectiveness of \emph{MementoMaps} of varying sizes by measuring the \emph{Accuracy} using 3.3M unique \emph{URIs} from \emph{MemGator} logs.
We found that a \emph{MementoMap} of less than 1.5\% \emph{Relative Cost} can correctly identify the presence or absence of 60\% of the lookup \emph{URIs} in the corresponding archive without any false negatives.
We open-sourced our implementation code under a permissive license~\cite{mementomap:gh}.
For dissemination and discovery of \emph{MementoMaps} we proposed the ``\texttt{mementomap}'' \emph{well-known URI} suffix and the ``\texttt{mementomap}'' \emph{link relation}.\looseness=-1

The trend shown in Figure~\ref{img:reduction-rates} opens up many possibilities to try, such as, to fit them as Heaps' Law~\cite{heapslaw} curves and estimate $K$ and $\beta$ parameters to then automatically identify the best roll up possibilities instead of asking a human to provide weights and supply other parameters.
The \emph{MementoMap} format proposed in this paper supports the ability to highlight inactive sub-trees within an active tree by being more specific, which will reduce false positives.
However, generating this information will require processing access logs or other out-of-band data sources.
Rolling the sub-tree up at \emph{H2} can be useful for large web archives and one way to explore this possibility is to identify globally less popular \emph{TLDs} that have a significant presence in an archive.
Currently, it is possible to do it manually, but not automatically.
A major goal of this work is to push for adoption of \emph{MementoMap} by adding out-of-the-box support in major archival replay systems.
We would also like to investigate the possibility of routing non-\emph{HTML} lookup requests by utilizing \emph{MementoMap} generated for \emph{HTML} mementos only.
The motivation comes from the assumption that page requisites are generally co-located with the parent page, hence we can leverage the information present in the \emph{Referer} header of embedded resources to identify potential archives to poll from.

\section{Acknowledgements}

This work is supported in part by National Science Foundation grant IIS-1526700.

\bibliographystyle{ACM-Reference-Format}
\bibliography{ref} 

\end{document}